\documentclass[allclo]{FBSart}
\usepackage{amsfonts}
\usepackage{amsmath}
\usepackage{amssymb}
\usepackage{graphicx}
\def\QED{\mbox{\rule[-1.5pt]{6pt}{10pt}}}
\def\re{{\rm Re\,}}
\def\Ran{{\rm Ran\,}}
\def\spect{{\rm spect\,}}

\def\R{\mathbb{R }}
\def\dsS{\mathbf{S }}

\def\HH{\mathcal{H }}
\def\OO{\mathcal{O }}

\title{On critical stability of three quantum charges interacting through delta potentials}

\author{H.D.\,Cornean\instnr{1}\thanks{\textit{E-mail address:} 
 cornean@math.aau.dk}, 
 P. Duclos\instnr{2} \thanks{\textit{E-mail address:} 
 duclos@univ-tln.fr}, 
 B. Ricaud
\instnr{2} \thanks{\textit{E-mail address:} 
 ricaud@cpt.univ-mrs.fr} }

\instlist{Dept. of Math., Aalborg University, Fredrik Bajers Vej 7G,  9220 Denmark

\and

Centre de Physique Th\'eorique UMR 6207 - Unit\'e Mixte de Recherche du
CNRS et des Universit\'es Aix-Marseille I, Aix-Marseille II et de
l' Universit\'e du Sud Toulon-Var - Laboratoire affili\'e \`a la
FRUMAM, Luminy Case 907, F-13288 Marseille Cedex 9 France}

\runningauthor{H. Cornean, P. Duclos, B. Ricaud}
\runningtitle{Critical stability of three quantum charges with delta self-interactions}
\begin{document}

\maketitle
\begin{abstract}
We consider three one dimensional quantum, charged and spinless
particles interacting through delta potentials. We derive  sufficient
conditions which guarantee the existence of at least one bound state.

\end{abstract}

%%%%%%%%%%%%%%%
\section{Introduction}%
%%%%%%%%%%%%%%
 Denote by $ x_i, m_i, Z_ie$,  $i=1,2,3$, the position, mass and
 charge of the $i$-th particle. Our system is formally described by the Hamiltonian
$
\sum_{i=1}^3-{\hbar^2\over 2m_i}\partial_{x_i}^2+\sum_{1\le i<j\le 3}Z_iZ_je^2\delta(x_i-x_j)
\quad\mbox{acting in $L^2(\R^3$)}
$ 
which is defined as the unique self-adjoint operator associated to the
quadratic form with domain $\HH^1(\R^3)$:
$$
\sum_{i=1}^3{\hbar^2\over2m_i}\|\partial_{x_i}\psi\|^2+\sum_{1\le
i<j\le3}Z_iZ_je^2\int_{x_i=x_j}|\psi(\sigma_{i,j})|^2d\sigma_{i,j},\quad \psi\in\HH^1(\R^3).
$$
Here $\sigma_{i,j}$ denotes a point in the plane $x_i=x_j$.  
We will consider the cases 
$
m_1=m_2=:m>0,\quad  m_3=:M>0\quad Z_1=Z_2=-1,\quad Z_3=:Z>0
$
and answer to the question: {\sl for what values of $m/M$ and $Z$ does
  this system possess at least one bound state after removing the
  center of the mass?} 

There is a huge amount of literature on 1-$d$ particles interacting through
delta potentials either all repulsive or all attractive, but rather
few papers deal with the mixed case. We mention the work of Rosenthal,
\cite{R},  where he considered $M=\infty$. The aim of this paper is to
make a systematic mathematical study of the Rosenthal results and
extend  them to the case $M<\infty$. It has been shown in \cite{BD}
and \cite{CDP} that these delta models serve as effective Hamiltonians
for atoms in intense magnetic fields or 
quasi-particles in carbon nanotubes. As one can see in (\cite{KB},
\cite{O}, \cite{SPMP},\cite{EI}), they also seem to be relevant for atomic wave guides, nano and leaky wires.

%%%%%%%%%%%%%%%%%%%%%%%%%%%%%
\section{The spectral problem}%
\begin{figure}
\begin{minipage}{60mm}
\begin{tabular}{|c|c|c|c|c|}
\hline
${M\over m}$ & $\theta_{2,3}$ &  $\theta_{1,2}$ &$\alpha^2$
&$\nu(\alpha)$\cr
\hline
$1$ & $2\pi/3$ & $2\pi/3$& $3/4$ & $1$  \cr
$\infty$ & $3\pi/4$ &$\pi/2$&$1/4$&$1/\sqrt{2}$ \cr
\hline
\end{tabular}
\end{minipage}
\begin{minipage}{60mm}
\centering
\includegraphics[width=5cm]{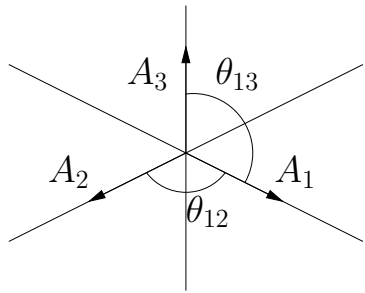}
\end{minipage}
\caption{Left: table with corresponding values of angles and masses, right: support of the delta potentials with the unit vectors
$A_i$'s.}\label{fig1}
\end{figure}

%%%%%%%%%%%%%%%%%%%%%%%%%%%%%%
%%%%%%%%%%%%%%%%%%%%%%
\noindent {\bf Removing the center of mass.}%
%%%%%%%%%%%%%%%%%%%%%%
 $\quad$ Using the Jacobi coordinates: $x:=x_2-x_1$, $y:= x_3-(m_1x_1+m_2x_2)/(m_1+m_2)$
and $z:=\sum_{i}m_ix_i/\sum_im_i$ we get the 2-$d$ relative motion
formal Hamiltonian 
$\widetilde H=
-{\hbar^2\over m}\partial_x^2-{2m+M\over
4mM}\hbar^2\partial_y^2+e^2\delta(x)-Ze^2\delta(y-{x\over2}) -Ze^2\delta(y+{x\over
2}).
$
Define $\alpha^2:=(M+2m)/4M$ and
$\nu(\alpha):=\sqrt{1/4+\alpha^2}$. Let $J$ be the Jacobian of 
the coordinate change $(x',y')=\{2\nu(\alpha)\hbar^2/(mZe^2)\}(x,\alpha
y)$, and define the unitary
$(U^{-1}f)(x,y)=\sqrt{J}f(x',y')$. Consider three unit vectors of
$\R^2$ given by 
$A_1:={1\over\nu(\alpha)}\left(\alpha,-{1\over2}\right)$, 
$A_2:={1\over\nu(\alpha)}\left(-\alpha,-{1\over2}\right)$, and 
$A_3:=(0,1)$. Define $A_i^\perp$ as $A_i$ rotated by $\pi/2$ in the positive sense.
Then $U\widetilde{H} U^{-1}=\{mZ^2
e^4\}/\{2\hbar^2\nu(\alpha)^2\}\; H$, where:
$$
H:=-{1\over 2}\partial_x^2-{1\over 2}\partial_y^2-\delta(A_1^\perp.(x,y))
-\delta(A_2^\perp. (x,y))+\lambda\delta(A_3^\perp . (x,y)),\quad \lambda:={\nu(\alpha)\over
  Z}.
$$
We denote by $\theta_{i,j}$ the angle between the vectors $A_i$ and
$A_j$. We give some typical values of all these parameters (see fig. \ref{fig1}).
\vspace{0.2cm}

%%%%%%%%%%%%%%
\noindent{\bf The skeleton}%
%%%%%%%%%%%%%%
 $\quad$ Let $A$ be unit vector in $\R^2$. If one introduce the "trace" operator  $\tau_{A}:\HH^1(\R^2)\to L^2(\R)$ defined as $(\tau_{A}\psi)(s):=\psi( sA)$ and if we let $\tau:\HH^1(\R^2)\to\bigoplus_{i=1}^3 L^2(\R)$ be defined as $\tau:=(\tau_{A_1},\tau_{A_2},\tau_{A_3})$,
we may rewrite the Hamiltonian $H$ as $H_0+\tau^\star g\tau$ where $2H_0$ stands for the free Laplacian and $g$ is the $3\times3$ diagonal matrix with entries $(-1,-1,\lambda)$. Denoting $R_0(z):=(H_0-z)^{-1}$and $R(z):=(H-z)^{-1}$ the resolvents of $H_0$ and $H$,  one derives at once, with the help of the second resolvent equation, the formula for any $z$ in the resolvent sets of $H_0$ and $H$:
\begin{align}\label{rezform}
R(z)=R_0(z)-R_0(z)\tau^*(g^{-1}+\tau R_0(z)\tau^*)^{-1}\tau
R_0(z).
\end{align}
Using the HVZ theorem (see \cite{S} for the case with form-bounded
interactions), we can easily compute the essential spectrum:
$\sigma_{{\rm ess}}(H)= [-{1\over2},\infty)$. Its bottom is given by the infimum of the spectrum of the subsystem made by the positive charge and one negative charge. 

From this and formula \eqref{rezform} it is standard to prove the
following lemma:
\vskip2mm
%%%%%%%%%%%%%%%%%%%
\noindent{\bf Lemma~1}.%
%%%%%%%%%%%%%%%%%%%
 $\quad$ {\it Let $k>{1\over\sqrt{2}}$. Define $\dsS:=  k\,g^{-1}+\tau
R_0(-1)\tau^*$. Then $E=-k^{2}<-{1\over2}$ is a discrete eigenvalue
of $H$ if and only if
$
\ker(  g^{-1}+\tau R_0(E)\tau^*)\ne\{0\}$. Note that up to a scaling
this is the same as $\ker\dsS\ne\{0\}$. Moreover, ${\rm mult}(E)={\rm dim}(\ker\dsS)$.}
\vskip2mm

The spectral analysis is thus reduced to the study of $\dsS$, a
$3\times 3$ matrix of integral operators each acting in $L^2(\R)$.  We call $\dsS$ the {\sl skeleton} of $H$.
Let us denote by $T_{A,B}:=\tau_AR_0(-1)\tau_B^\star$,
$T_0:=T_{A,A}$, by $\theta_{A,B}$  the angle between 
two unit vectors $A$ and $B$,  and by $\widehat T_{A,B}$ the Fourier image of $T_{A,B}$.  Then the kernel of $\widehat T_{A,B}$  when
$\theta_{A,B}\not\in \{0,\pi\}$, and of $\widehat T_0$ read as:
\begin{equation}\label{Ttheta}
\widehat T_{A,B}(p,q)=
{1\over2\pi|\sin(\theta_{A,B})|}{1\over
\left({p^2-2\cos(\theta_{A,B}) pq +q^2\over 2\sin^2(\theta_{A,B})}+1\right)},\quad
\widehat T_0(p,q)={\delta(p-q)\over\sqrt{p^2+2}}.
\end{equation}
Then $\widehat T_0$ is a bounded multiplication operator, and
$\widehat T_{A,B}$ only depends on $|\theta_{A,B}|$. Consequently we
denote in the sequel $T_{A_i,A_j}$ by
$T_{\theta_{i,j}}$ or $T_{i,j}$  .

\vspace{0.2cm}

%%%%%%%%%%%%%%
\noindent{\bf Reduction by symmetry.}%
%%%%%%%%%%%%%%
 $\quad$ 
$H$ and $\dsS$ enjoy various symmetry properties which follow from the fact that two particles are identical. Let 
$
\pi:L^2(\R)\to L^2(\R)
$
be the parity operator, i.e. $\{\pi\varphi\}(p)=\varphi(-p)$ and
denote by $\pi_1:=\pi\otimes 1$ and $\pi_2:=1\otimes\pi$ the 
unitary symmetries with respect to the $x$ and $y$ axis. One verifies
that for all $i,j\in\{1,2\}$, we have $[\pi_i,H]=0$ and
$[\pi_i,\pi_j]=0$. 
Thus if we denote by $\pi_i^\alpha$, $\alpha=+,-$ the eigenprojectors of $\pi_i$ on the even, respectively odd functions we may decompose $H$ into the direct sum
$$
H=\bigoplus_{\alpha\in\{\pm\},\,\beta\in\{\pm\}} H^{\alpha,\beta},\quad  H^{\alpha,\beta}:=\pi_1^\alpha\pi_2^\beta H.
$$
Similarly let $\Pi,\sigma:L^2(\R^3)\to L^2(\R^3)$ defined by
$(\Pi\psi)(-p):=\psi(-p)$ and
$\sigma\psi=\sigma(\psi_1,\psi_2,\psi_3):=(\psi_2,\psi_1,\psi_3)$.
They both commute with $\dsS$, 
and also $[\Pi,\sigma]=0$. 

Let $\Pi^\alpha$ and $\sigma^\alpha$, $\alpha=+,-$, denote the eigenprojectors of $\Pi$ and $\sigma$ symmetric and antisymmetric resp..
 Then we can write $
\dsS=\bigoplus_{\alpha\in\{\pm\},\,\beta\in\{\pm\}} \dsS^{\alpha,\beta},\quad  \dsS^{\alpha,\beta}:=
\Pi^\alpha\sigma^\beta\dsS$. 
 
 From the expression of $\widehat T_\theta(p,q)$ one also sees that
 $[\pi,T_\theta]=0$ so that $T_\theta$ decomposes into
 $T_\theta^+\oplus T_\theta^-$ where $T_\theta^\pm:=\pi^\pm T_\theta$.   As
 usual we shall consider $T_\theta^\pm$ as operators acting in
 $L^2(\R_+)$. 
Due to these symmetry properties we have
$\ker\dsS=\bigoplus_{\alpha,\beta}\ker\dsS^{\alpha,\beta}$, and each
individual null-space can be expressed as the null-space of a single
operator acting in $L^2(\R_+)$ that we call {\sl effective skeleton}.
We gather in the following table the four effective skeletons we have  to consider with their corresponding subspaces in $L^2(\R^2)$:

\vskip2mm
\begin{minipage}{120mm}
\begin{center}
\begin{tabular}{|c|c|c|}
\hline
$S^{\alpha,\beta}$& effective skeleton & subspace in $L^2(\R^2)$\cr
\hline
$++$    & $k-T_0-T_{1,2}^++2T_{2,3}^+(T_0+k\lambda^{-1})^{-1}T_{2,3}^+$ & $\Ran\pi_1^+\pi_2^+$\cr
$-+$    & $k-T_0-T_{1,2}^-+2T_{2,3}^-(T_0+k\lambda^{-1})^{-1}T_{2,3}^-$ & $\Ran\pi_1^+\pi_2^-$\cr
$+-$    & $k-T_0+T_{1,2}^+$                                   & $\Ran\pi_1^-\pi_2^-$\cr
$--$    & $k-T_0+T_{1,2}^-$                                   & $\Ran\pi_1^-\pi_2^+$\cr
\hline 
\end{tabular}
\vskip1mm
Table~1.
\end{center}
\end{minipage}

%%%%%%%%%%%%%%%%%%%%
\section{Sectors without bound states}%

\vspace{0.2cm}
%%%%%%%%%%%%%%%%%%%%
%%%%%%%%%%%%%%%%%%%%%%%%%%
\noindent {\bf Properties of the $T_\theta$ operators.}%
%%%%%%%%%%%%%%%%%%%%%%%%%%
 $\quad$ From (\ref{Ttheta}) we get $0\le T_0\le 1/\sqrt{2}$. Then
 $T_\theta$ is self-adjoint and has a finite Hilbert-Schmidt norm. 
The proof of the following lemma in not at all obvious, but will be omitted due to the lack of space.
\vskip2mm
%%%%%%%%%%%%%
\vspace{0.2cm}
\noindent{\bf Lemma~2}.%
%%%%%%%%%%%%%
{\sl\  For all $\theta\in[\pi/2,\pi)$ one has $\pm T_\theta^\pm \ge0$
  and the mapping $[\pi/2,\pi)\ni \theta\mapsto\pm\inf T_\theta^\pm$
  is strictly increasing.}

\vspace{0.2cm}
%%%%%%%%%%%%%%%%%%%%
%%%%%%%%%%%%%%%%%%%%%%%%%%
\noindent {\bf Absence of bound state in the odd sector with respect to $y$.}%
%%%%%%%%%%%%%%%%%%%%%%%%%%
 $\quad$ We have the following result: 

\vspace{0.2cm}

\noindent{\bf Theorem~3}.%
%%%%%%%%%%%%%
{\sl\  For all $Z>0$ and all $0< M/m\le \infty$, $H$ has no bound
  state in the symmetry sector $\Ran\pi_2^-$.}

\noindent {\it Proof.} The symmetry sector $\Ran\pi_2^-$ corresponds
to the second and third lines in Table~1. For the third line 
one uses that $T_{1,2}^+\ge0$ by Lemma~2, and that $k>1/\sqrt{2}$ since we are
looking for eigenvalues below $\sigma_{\rm ess}(H)=[-{1\over
2},\infty)$. Hence 
$$
k-T_0+T_{1,2}^+\ge k-{1\over\sqrt{2}}>0
$$ 
thus $\ker(k-T_0+T_{1,2}^+)=\{0\}$, and by Lemma~1 this shows that $H$ has no
eigenvalues in $\Ran\pi_1^-\pi_2^-$. By the same type of arguments one has:
$
k-T_0-T_{1,2}^-+2T_{2,3}^-(T_0+k\lambda^{-1})^{-1}T_{2,3}^-\ge k-{1\over\sqrt{2}}>0. \ \QED
$
\vskip2mm
%%%%%%%%%%%%%
\noindent{\bf Remark~4}.%
%%%%%%%%%%%%%
$\quad $ The above theorem has a simple physical interpretation. Wave functions which are antisymmetric in the $y$ variable are those for which the positive charge has a zero probability to be in the middle of the segment joining the negative charges. A situation which is obviously not favorable for having a bound state.

\vspace{0.2cm}
%%%%%%%%%%%%%%%%%%%%
%%%%%%%%%%%%%%%%%%%%%%%%%%
\noindent {\bf Absence of bound state in the odd-even sector with
  respect to $x$ and $y$.}%
%%%%%%%%%%%%%%%%%%%%%%%%%%
 $\quad$ \noindent Looking at the fourth line of Table~1 we have to consider
\begin{equation}\label{Smoinsmoins}
S^{-,-}(k):=k-T_0+T_{1,2}^-=:
\sqrt{k-T_0}\left(1+\widetilde T_{1,2}^-(k)\right)\sqrt{k-T_0}
\end{equation}
where $\widetilde
T_{1,2}^-(k):=(k-T_0)^{-{1\over2}}T_{1,2}^-(k-T_0)^{-{1\over2}}$. Here
we will only consider the case $M\ge m$, i.e. $\pi/2\le \theta_{1,2}\le 2\pi/3$. Assume that we can prove that
$
\widetilde
T_{1,2}^-(2^{-{1\over2}})\ge -1
$
for $\theta_{1,2}=2\pi/3$, this will imply that $S^{-,-}(2^{-{1\over2}})\ge0$ first for $\theta_{1,2}=2\pi/3$ and then for all $\pi/2\le \theta_{1,2}\le 2\pi/3$ by the monotonicity of $\inf T_{1,2}^-$ with respect to $\theta$ as stated in Lemma~2; finally looking at (\ref{Smoinsmoins}) this will show that $S^{-,-}(k)>0$ for all $k>1/\sqrt{2}$ and therefore that $\ker S^{-,-}(k)=\{0\}$.
But 
$
\widetilde
T_{1,2}^-:=T_{1,2}^-(2^{-{1\over2}})
$ ( for $ \theta_{1,2}=2\pi/3$) 
 is Hilbert-Schmidt  since its kernel decay at infinity faster than
 the one of $T_{1,2}^-$ and it has the following 
behavior at  the origin:
$
\widetilde
T_{1,2}^-(p,q)\sim
-{16\sqrt{2}\over3\sqrt{3}\pi}+\OO\left((p^2+q^2)\right).
$
It turns out that 
$-1$ is an eigenvalue of $\widetilde
T_{1,2}^-$ with eigenvector  
$$
\R_+\ni p\mapsto \frac{
\left [\frac{1}{\sqrt{2}}-\frac{1}{\sqrt{p^2+2}}\right ]^{1/2}}{p\left(2 p^2+3
\right) }\quad\buildrel p\to0\over\sim\quad
\frac{1}{3\,2^{\frac{5}{4}}}+\OO(p^2)
$$
and since the Hilbert-Schmidt norm of $\widetilde T_{1,2}$ can be evaluated numerically to $\|\widetilde T_{1,2}^-\|_{HS}\le 1.02$, all the other eigenvalues of $\widetilde T_{1,2}^-$ are above $-1$. Thus we have proved the 
\vskip2mm
%%%%%%%%%%%%%%
\noindent{\bf Theorem~5}.%
%%%%%%%%%%%%%%
{\sl\  For all $Z>0$ and all $1\le  M/m\le \infty$, $H$ has no bound state in the symmetry sector $\Ran\pi_1^-\pi_2^-$.}

%%%%%%%%%%%%%%%%%%%
\section{The fully symmetric sector}%
%%%%%%%%%%%%%%%%%%%
According to Table~1, we need to find under which conditions one has $\ker S^{+,+}(k)\ne\{0\}$ where 
$$ 
S^{+,+}(k):=k-T_0-K(k),\quad{\rm with}\quad K(k):=T_{1,2}^+-2T_{2,3}^+(T_0+k\lambda^{-1})^{-1}T_{2,3}^+.
$$ 
The proof of the following lemma is an easy application of Fredholm 
and analytic perturbation theory:

\vskip2mm
%%%%%%%%%%%%%
\noindent{\bf Lemma~6}.%
%%%%%%%%%%%%%
{\it  {\rm (i)} $\{\re k^2>0\}\ni k\mapsto S^{+,+}(k)$ is a bounded analytic self-adjoint family
of  operators.

\noindent{\rm (ii)} If $\inf \sigma\left (S^{+,+}(2^{-{1\over2}})\right )<0$, then there exists  $k>1/\sqrt{2}$ so that
$\ker S^{+,+}(k)\ne \{0\}$.
}

\vskip2mm

Denote by $\mathcal{K}(p,q)$ the integral kernel of
$K(2^{-{1\over2}})$. Our last result is:
\vskip2mm
%%%%%%%%%%%%%%%%%%%%%
\noindent{\bf Theorem~7}.%
%%%%%%%%%%%%%%%%%%%%%
{\sl\ For all  $0< M/m\le \infty$, $H$ has at least one bound state in the symmetry sector $\Ran\pi_1^+\pi_2^+$ if $Z$ is such that
$
\mathcal{K}(0,0)>0.
$
}
\vskip2mm

\noindent {\bf Proof.} We will now look for an upper bound on $\inf S^{+,+}(2^{-{1\over2}})$ by the variational method. Let 
$
j\in C_0^\infty(\R_+,\R_+)
$
so that $\int_{\R_+} j(x) dx=1$ and define two families of  functions: 
$
\forall \epsilon>0,\
\psi_\epsilon(p):=\epsilon^{-1}j(p\epsilon^{-1}),\
\phi_\epsilon := \frac{\sqrt{\epsilon}}{||j||}\psi_\epsilon,\
\|\phi_\epsilon\|=1.
$
We know that $\psi_\epsilon$ converges as $\epsilon\to0$ to the Dirac distribution. First one has
\begin{align}
((2^{-{1\over2}}-T_0)\phi_\epsilon,\phi_\epsilon)&=
{1\over\epsilon\sqrt{2}\|j\|^2}\int_{\R_+} 
%:
[1-(1+p^2/2)^{-1/2}]j^2({p/\epsilon})dp\nonumber \\
&\le{\epsilon^2\over 2\sqrt{2}\|j\|^2}\int_{\R_+} p^2j(p)^2dp.\end{align}
Then 
one has
$
(K(2^{-{1\over2}})\phi_\epsilon,\phi_\epsilon)={\epsilon\over\|j\|^2}(K(2^{-{1\over2}})\psi_\epsilon,\psi_\epsilon)=
{\epsilon\over\|j\|^2}\left(\mathcal{K}(0,0)+\OO(\epsilon)\right)
$
so that
$(S^{+,+}(2^{-{1\over2}})\phi_\epsilon,\phi_\epsilon)=2^{-{1\over2}}-(T_0\phi_\epsilon,\phi_\epsilon)-(K(2^{-{1\over2}})\phi_\epsilon,\phi_\epsilon)$
will be negative for $\epsilon>0$ small enough, provided
$\mathcal{K}(0,0)>0$. \QED

\vspace{0.2cm}

 It is possible to compute $\mathcal{K}(0,0)$
analytically. It can be shown that there exists $Z_c^{\rm ub}(M/m)$
such that for any $Z$ larger than this value, we have
$\mathcal{K}(0,0)>0$. If we now define the critical $Z$ as $
Z_c(M/m):=\inf \{ Z>0,\, H=H(Z,M/m)\  \mbox{has at least one bound state}\},
$ it follows from our last theorem that $Z_c(M/m)\leq Z_c^{\rm ub}(M/m)$.

\begin{figure}
\centering
\includegraphics[width=8cm]{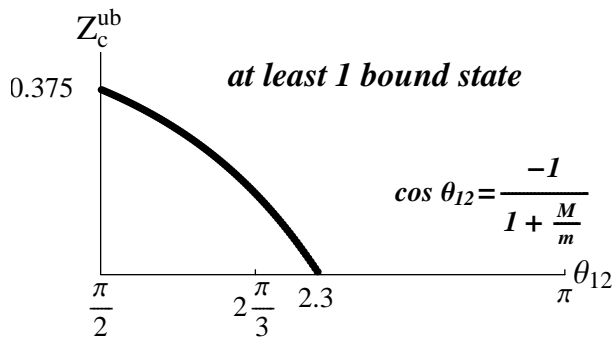}
\caption{Graph of $Z_c^{\rm ub}$.}\label{fig2}
\end{figure}

The curve $Z_c^{\rm ub}(M/m)$ is plotted on figure~\ref{fig2}, where
we used $\theta_{1,2}$ instead of the ratio $M/m$.
\vskip2mm
%%%%%%%%%%%%%%%%%%%%%
\noindent{\bf Remarks~8}.%
%%%%%%%%%%%%%%%%%%%%%
\ (a) Rosenthal found numerically $Z_c^{\rm ub}(\frac{\pi}{2})$, i.e. $Z_c^{\rm ub}$ for $M=\infty$ to be $0.374903$. With our analytical expression of $\mathcal{K}(0,0)$ we know this value to any arbitrary accuracy: $Z_c^{\rm ub}(\frac{\pi}{2})=0.37490347747000593278...$
\vskip1mm
\noindent (b) The above curve shows that an arbitrarily small positive charge of mass $M<0.48m$ can bind two electrons. However we believe that the exact critical curve will show that $M<m$ and $Z>0$ is sufficient to bind these two electrons.

\begin{acknowledge}
H.C. acknowledges support from the Statens Naturvidenskabelige
Forskningsr{\aa}d grant {\it Topics in Rigorous
  Mathematical Physics}, and partial support through the European Union's IHP
network Analysis $\&$ Quantum HPRN-CT-2002-00277. P.D. and B.R. thanks the Universit\'e du Sud Toulon-Var for its support from its mobility program.
\end{acknowledge}

\end{document}